\documentclass[]{aa}
\usepackage{graphicx}
\begin{document}
\thesaurus{11.01.1; 11.03.1; 11.03.2; 11.03.4; 11.04.1; 11.04.2; 11.05.2;
11.09.5; 12.12.1}
\title{On the star-formation properties of emission-line galaxies in and 
around voids
\thanks{Based on observations obtained at the German-Spanish Observatory at
Calar Alto, Almeria, Spain}}
\author{Cristina C. Popescu \inst{1,2}
\and  Ulrich Hopp  \inst{3}
\and Michael R. Rosa \inst{4}\thanks{Affiliated to Astrophysics Division of 
the Space Science Department of the European Space Agency}
}

\offprints{Cristina C. Popescu (email address: popescu@levi.mpi-hd.mpg.de)}
\institute{Max Planck Institut f\"ur Kernphysik, Saupfercheckweg 1, 
           D--69117 Heidelberg, Germany
\and The Astronomical Institute of the Romanian Academy, Str. Cu\c titul de
Argint 5, 75212, Bucharest, Romania
\and Universit\"atssternwarte M\"unchen, Scheiner Str.1, 
       D--81679 M\"unchen, Germany
\and The Space Telescope European Coordinating Facility, European Southern
Observatory, Karl-Schwarzschild-Stra{\ss}e 2, 85740 Garching, Germany}
\date{Received July 8; accepted: August 27, 1999}
\maketitle
\markboth{C.C.Popescu et al.}{On the star-formation properties of 
emission-line galaxies in and around voids}
\begin{abstract}
We present a study of the star formation properties of a sample of emission
line galaxies (ELGs) with respect to their environment. This study is part of
a bigger project that aimed to find galaxies in voids and to investigate the
large scale structure of the ELGs. A survey for ELGs was therefore conducted
with the result that 16 galaxies have been found in very low density
environments, of which 8 ELGs were found in two very well
defined nearby voids. The sample presented here contains some galaxies
identified in voids, as well as in the field environment that delimited the 
voids. These
 ELGs are all Blue Compact Galaxies (BCGs), and all void galaxies are also 
dwarfs. Both void and field sample contain the same mixture of morphological
subtypes of BCDs, from the extreme Searle-Sargent galaxies to the 
Dwarf-Amorphous Nuclear-Starburst galaxies. The main result of this study is 
that field and void galaxies seem to 
have similar star formation rates (SFR), similar ratios between the current 
SFR and their average past SFR and similar mean SFR surface densities. There 
is no trend in metallicity, in the sense that void galaxies would have 
lower metallicities than their field 
counterparts. The field-cluster dichotomy is also discussed using available
results from the literature, since our sample does not cover the cluster
environment. The implication of our findings are discussed in connection with
the theories for the formation and evolution of BCDs. 
\end{abstract}

\keywords{galaxies: compact - galaxies: dwarf - galaxies: irregular - 
galaxies: evolution -  galaxies: abundances -  galaxies: clusters: general - 
galaxies: clusters: individual: Virgo Cluster - galaxies - 
distances and redshifts - large scale structure of Universe}

\section{Introduction}

Popescu et al. (1996, hereafter PHHE96, and 1998, hereafter PHHE98) conducted 
a survey for emission-line galaxies (ELGs) towards nearby voids, with the aim 
of finding faint galaxies within the voids. This was motivated by the 
question of whether the void regions represent real structures, or whether they merely 
appear as such due to selection effects. 
According to some theories of galaxy formation 
(Dekel \& Silk 1986), that were worked out in the frame of the cold dark 
matter (CDM) scenarios including biasing, dwarf 
galaxies should originate from the 1\,${\sigma}$ fluctuations, and ought to be 
more evenly distributed than the high rare density peaks 
which form the giants. In these scenarios the dwarf galaxies should trace the 
underlying dark matter and are expected to fill the voids. 
Since there was a hint that emission-line galaxies are more evenly 
distributed (Salzer 1989, Rosenberg et al. 1994), this kind of objects were
chosen to be searched, and primarily dwarf HII galaxies or BCDs. In passing 
we note that these two terms are used in the literature interchangeably for 
the same physical 
type of object. While the name of 
BCD was used for objects that were classified on morphological criteria 
(Binggeli et al. 1985 - for the Virgo Cluster Catalog), the term of HII 
galaxy was introduced for objects discovered on spectroscopic surveys for 
emission line galaxies (Sargent and Searle 1970, Searle and Sargent 1972). 

The survey of PHHE96,PHHE98 was based on a sample selected from the 
Hamburg Quasar Survey (HQS) (Hagen et al. 1995) - IIIa-J digitised objective 
prism plates and the main selection criterion was the presence of 
emission-lines, mainly of the [OIII]\,${\lambda}$5007 line. The result of the 
survey is that no homogeneous void 
population was found (Popescu et al. 1997). Nevertheless 16 ELGs
were identified in very low density environments, from which at 
least 8 ELGs were found in two very well defined nearby voids. All the 
isolated galaxies were dwarf galaxies (M$_B < -18.5$, diameters 
${\rm D}< 10$\,kpc) of BCD class. 
In this paper we compare the spectroscopic properties of 11 isolated ELGs 
from the void environment (hereafter Void) with the spectroscopic properties of
44 galaxies from the field environment (hereafter Field), in order to 
evaluate the influence of 
the environment on the evolution of dwarf HII galaxies. 

The isolated galaxies as well as those galaxies found in the 
filaments and sheets surrounding the voids belong to the same 
sample and thus they were selected with the same selection criteria. Then 
the two subsamples do not suffer from different selection biases and 
therefore are valid for studies of environmental effects. Furthermore, 
the objects were all observed with the same telescope and set-ups, providing 
us with a homogeneous set of data. Popescu et al. (1997) have also shown that 
the sample used for the statistical analysis of the large-scale structure is 
a complete sample. Both subsamples contain the same type of objects, with the 
same morphological mixture. If, however, different 
subsamples containing varying percentage of morphological types are 
compared to infer environmental influences, then any difference found 
between subsamples will instead reflect the differences between different 
Hubble types. It could well be that this effect is the main reason behind
the differing answers of different studies on environmental effects. 

The influence of the environment on the mechanism that
control star formation is likely one of the most important aspects in 
understanding the
origin and evolution of galaxies. It is commonly believed that dwarf galaxies 
are heavily influenced by their environment. While many studies 
concentrated on the differences between cluster and field galaxies, very 
little is known about differences between Field and Void galaxies. This is 
especially true since voids appear to be real empty environments, while only 
a few galaxies were identified in these regions. Therefore the statistics of
void galaxies is very poor. Much effort was put in the study of the Bootes
void, a huge low density region first discovered by Kirshner et al. (1981)(see
also Kirshner et al. 1983a,b). Up to now 58 galaxies were identified in this
void (Szomoru et al. 1996). But Bootes void is beyond the distance at which 
the large scale
structure is well defined by the present catalogues. Therefore it is not clear
whether the galaxies found in this void are real void galaxies, or whether they
are only field galaxies that were not well sampled by the present magnitude
limited surveys. Weistrop et al. (1995) showed that the void galaxies in Bootes
were luminous galaxies, not the low mass, low surface brightness galaxies
predicted to be found in voids. We therefore especially selected nearby voids
which are very well defined in the distribution of normal galaxies.

Because our sample of Void galaxies contains only 11 objects we will not try
to search for statistical differences in the shape of the distributions. Thus
no statistical tests (e.g. the  Kolmogorov-Smirnov test) will be applied. Rather
 we will investigate whether the distribution of different
spectroscopic parameters of Void galaxies occupies the same locus in 
diagrams/histograms as in the case of Field galaxies. We will show that there
do not appear to be any differences between Void and Field galaxies under such
a comparison. 

The paper is organised as follows. In section 2 we describe the sample used for
the study of the environmental effects, in section 3 we compare the star
formation properties of Field and Void galaxies and in section 4 we study the
metallicities of our program galaxies with respect to their environment. In 
section 5 we investigate the dichotomy Field - Cluster, with direct 
application to the galaxies in the Virgo Cluster core  and in  section 6 we 
discuss our findings and we give a summary of the results.

\section{The sample}

In this paper we study a subsample of the ELGs found in the survey  of
PHHE96,PHHE98, namely galaxies in a region North to the \lq\lq 
Slice of the Universe\rq\rq\ (de Lapparent et al. 1986). In order to describe
the large-scale structure of this region we used 
a comparison catalogue of giant normal galaxies, namely the ZCAT 
(Huchra et al. 1992, Huchra et al. 1995). The comparison catalogue should 
trace the main structures in the nearby Universe and should properly define 
the nearby voids. The region under study is dominated by the field galaxies 
that belong to the \lq\lq Great Wall\rq\rq\ (Ramella, Geller and Huchra 1992), 
at a distance of 
$6500\,$km/s at ${\alpha}=12^{\rm h}00^{\rm m}$ and $9000\,$km/s at 
${\alpha}=15^{\rm h}30^{\rm m}$. At velocities less than 7500\,km/s, there are
more field galaxies, remnants of the \lq\lq Harvard Sticky Man\rq\rq\, but
without the Coma Cluster included. The field galaxies define at least two 
nearby voids (Popescu et al. 1997), one centred at
${\alpha}\sim13^{\rm h}15^{\rm m}$, ${\rm v}\sim3000\,$km/s, and the other one 
being just in front of the Great Wall. Beyond the Great Wall there is an
indication of more voids, but their sizes are much less constrained by the
galaxy distribution. 

We restrict our analysis on ELGs with velocities less
than 10000\,km/s, since for larger distances the comparison catalog quickly
thins out, and therefore we refrain from drawing conclusions about the reality
of voids beyond 10000\,km/s. The criteria for separating 
the sample in Field and Void are given in Popescu et al. (1997), based on the 
calculation of the nearest neighbour distances (D$_{NN}$). The D$_{NN}$ 
represent real separations in the 3-dimensional space and are calculated
(Popescu et al. 1997) as the separation between each ELG and its nearest 
neighbour from the comparison catalogue, for galaxies with blue magnitudes 
brighter than 15.5. The nearest neighbour distances of the void galaxies and
their clustering properties can be found in Table 3 from Popescu et
al. (1997). As an example, we illustrate the case of the void galaxy
HS1236+3937 (z=0.0184). It has its nearest neighbour ZCAT galaxy at a distance
8.68\,h$^{-1}$\,Mpc. At this redshift the average nearest 
neighbour distances between the field ELGs are 0.75\,h$^{-1}$\,Mpc. Thus the
galaxy HS1236+3937 differs from its field counterparts by a  factor of 12 
in its clustering properties. Furthermore, the 
mean nearest neighbour distance for the Void ELGs is 4.5\,h$^{-1}$\,Mpc, as 
compared to the mean D$_{NN}=1\,{\rm h}^{-1}$\,Mpc for the Field ELGs . The 
difference in the clustering properties of the two samples proves that 
they form two distinct populations.

The sample of ELGs, selected as described in the previous paragraph, was 
further constrained to contain only 
galaxies with reliable
spectrophotometry. A list of the selected sample galaxies
is given in Table 1, together with their environment: Field (F) or Void (V)
 galaxy (Column 9).  

The spectroscopic parameters of the ELGs (emission-line ratios, equivalent
widths and fluxes) together with a detailed description of the data is given
by Popescu and Hopp (1999), while the photometric parameters (B and R magnitudes,
diameters, morphologies and surface brightness profiles) are given by Vennik et
al. (1999). In Table 1 we list only the main parameters of the selected 
galaxies: absolute B magnitudes (Column 2), redshifts (Column 3), and the 
measured equivalent widths of H${\beta}$ (Column 4).

All ELGs included in this study were Blue Compact galaxies, and most of them
were also dwarfs. The very luminous Starburst Nucleus Galaxies (SBN) were 
not considered in this sample, 
the same being true for galaxies with an active nucleus. The BCGs were further
 classified (Vennik et al. 1999) into the morphological classes proposed by 
Salzer et al. (1989a,b), namely  Dwarf Amorphous Nuclear Starburst Galaxies 
(DANS), HII Hotspot Galaxies (HIIH), Dwarf HII Hotspot Galaxies (DHIIH),
Sargent-Searle Objects (SS), and Interacting Pairs (IP). The classification was
done based only on the photometric parameters (absolute magnitude, diameters)
and the morphological appearance on the CCD images, as described by Vennik et
al. (1999). Nevertheless, an independent check was done using the line ratios
of the emission-lines and their location in the diagnostic diagrams, and there
is an overall good agreement between the morphological and spectroscopic
classification. The type of each
galaxy is given in Table 1, Column (10), while their distribution is plotted in
Fig. 1.

\begin{figure}[htp]
\includegraphics[scale=0.4]{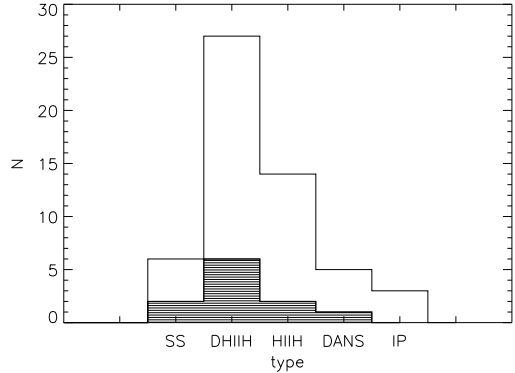}
\caption{The distribution of morphological types. The 
solid histogram is the distribution of the whole sample, while the hashed 
histogram is for Void galaxies.}
\end{figure} 

The distribution of morphological types for Void galaxies is also plotted with
hashed-histogram, indicating that
galaxies in Voids are spread over all the BCG subtypes. Thus the
two samples of Field and Void galaxies consist of the same morphological
types; accordingly any differences between the BCGs in these two samples is not due to a different mixture of 
classes.

\section{Star formation properties}


We first present the histogram of the H${\beta}$ equivalent widths (Fig. 2), 
which are considered to measure the strength and the age of the star-burst. Large values
of the W(H${\beta}$), which means stronger and more recent star-burst are
common in both the Field and the Void sample. The same is valid for the low
values of the W(H${\beta}$).

\begin{figure}[htp]
\includegraphics[scale=0.4]{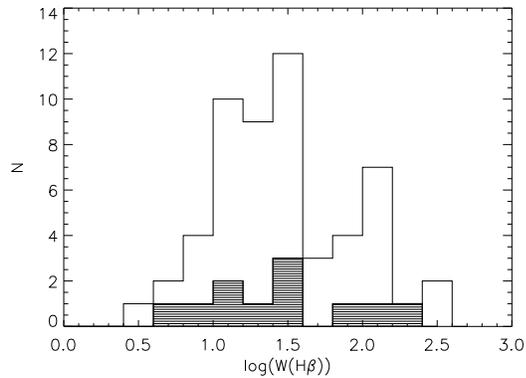}
\caption{The distribution of the H${\beta}$ equivalent widths (in units of 
$\AA$). The solid 
histogram is the distribution of the whole sample,
while the hashed histogram is for Void galaxies.}
\end{figure}


Further we use the luminosity of H$\beta$ to estimate the global current
star formation rates of the ELGs (Kennicutt 1983). For all the galaxies 
included in this study
the fluxes were measured with a 4$^{\prime\prime}$ slit width. For the small 
projected 
sizes of our dwarf galaxies, such an aperture is large enough to encompass 
most of their line emission. Thus we can expect that the H$\beta$ fluxes 
represent an accurate measurement of the total integrated H$\beta$ fluxes of 
the galaxies.

\begin{figure}[htp]
\includegraphics[scale=0.4]{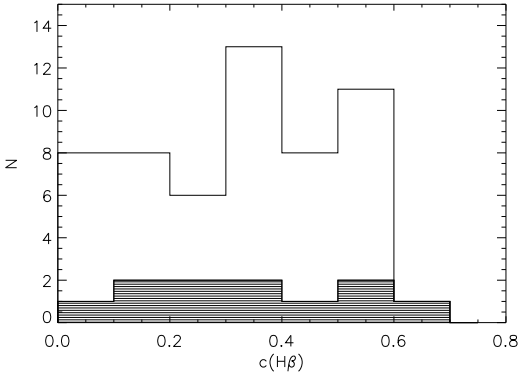}
\caption{The distribution of the extinction coefficient c(H$\beta$). The 
solid histogram is the distribution of the whole sample, while the hashed 
histogram is for Void galaxies.}
\end{figure} 
For all the galaxies with a strong underlying continuum emission (relative to
the  H$\beta$ line emission, W(H$\beta) < 20$\,$\AA$) we corrected the 
H$\beta$ fluxes for underlying stellar absorption with an assumed constant
equivalent width of 2\,$\AA$ (McCall, Rybski and Shields 1985). The H$\beta$ 
fluxes were corrected for reddening due to dust in our Galaxy and on the
galaxy being observed. We used the reddening coefficient c(H$\beta$), 
derived from the observed H$\alpha$/H$\beta$ Balmer line ratios. For the
extinction in our Galaxy we used the standard Galactic reddening law 
(Whitford 1958) and the extinction values from Burstein and
Heiles (1984). Afterwards we corrected for internal extinction making the 
reasonable assumption that the intrinsic Balmer-line ratios are equal to the
case B recombination values of Brocklehurst (1971), for an electron 
temperature of 10$^4$\,K and an electron density of 100\,cm$^{-3}$.  We used 
the LMC reddening law (Howarth 1983), reckoning that it should be closer to the
average reddening law for the dwarf galaxies of our sample than the standard
galactic one.   
The absorption coefficient c(H$\beta$) is listed for each galaxy in Table 1, 
Column 5. The H$ \beta$ luminosities are computed using a Hubble constant
${\rm H0}=75$\,km/s/Mpc and are listed in Table 1, Column 6. 
In Fig. 3 we give the distribution of the absorption coefficient c(H$\beta$)
for the whole sample as well as for the Void galaxies. Galaxies with differing 
amounts of internal extinction are spread evenly amongst
the Field and Void samples. 

\begin{figure}[htp]
\includegraphics[scale=0.5]{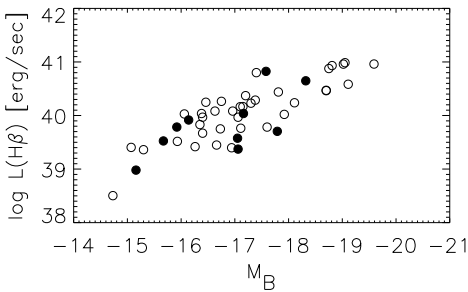}
\caption{Logarithm of H${\beta}$ luminosity as a function of 
blue absolute magnitude. Open circles are Field galaxies and filled circles are
Void galaxies.}
\end{figure}
Fig. 4 shows the H${\beta}$ luminosities against the blue absolute magnitude. The
open dots represent the Field galaxies while the filled dots the Void
galaxies. As expected, there is a good correlation between the two
quantities, and this correlation is present for both Field and Void galaxies.
Our Void galaxies seem to occupy  the same locus in the diagram as the Field 
objects,
except for the fact that the upper part of the correlation (large blue and 
H${\beta}$
luminosities) is not populated by the Void galaxies. This is once again
corroborating evidence (see Popescu et al. 1997) that all our Void galaxies 
were dwarfs. 

The H$\beta$ fluxes were then converted to total SFRs using the calibration of
Kennicutt (1998, see also Kennicutt et al. 1994), namely 
\begin{eqnarray}
{\rm SFR}({\rm M}_{\odot}\,{\rm yr}^{-1}) = \frac {2.85\times 
{\rm L}({\rm H}{\beta})}{1.26\times 10^{41}{\rm erg}\,{\rm s}^{-1}},
\end{eqnarray}
again in the case B recombination value of the intrinsic  
H$\alpha$/H$\beta$ Balmer ratio.

The IMF used in this conversion is a Salpeter function dN(m)/dm =-2.35 over 
the mass range ${\rm m}=0.1-100\,{\rm M}_{\odot}$. Kennicutt et al. (1998) 
indicated that 
the extended Miller-Scalo IMF used in the calibration of Kennicutt (1983) 
would produce nearly identical SFRs (only $8\%$ lower). The total SFRs are 
listed in Table 
1, Column 7. The  values of the L(H$\beta$) range between 
$3.19\times 10^{38} \le {\rm L}({\rm H}\beta) \le 9.69\times10^{40}$\,erg/sec, 
which is equivalent with a range in
star formation rates between 0.0072\,M$_{\odot}$/yr and 2.19\,M$_{\odot}$/yr.

Despite the fact that there is a clear correlation between M$_B$ and the 
H$\beta$ luminosity (Fig 4), L(H$\beta$) at any given M$_B$ has a scatter of
about one order of magnitude. If the total blue magnitude is a good measure 
of the average past star formation, then the scatter indicates an intrinsic 
variation in the ratio between the present star formation rate and the 
average past.

\begin{table*}
\begin{tabular}{cccrcccccl}
\multicolumn{10}{c}{{\bf Table 1 }}\\
\multicolumn{10}{c}{}\\
\hline\hline
& & & & & & & & & \\
(1) & (2) & (3) & (4) & (5) & (6) & (7) & (8) & (9) & (10) \\
&  & & & & & & & &\\
Galaxy & M$_B$ & z & W(H$\beta$) & c(H$\beta$) & L(H$\beta$) 
&SFR & log ${\Sigma}_{SFR}$ &   & type \\
       &       &   & [$\AA$]     &             & [erg/sec]   
&[M$_{\odot}\,{\rm yr}^{-1}$]      & 
[M$_{\odot}\,{\rm yr}^{-1}\,{\rm kpc}^{-2}$]  & \\
& & & & & & & & & \\
\hline
& & &  & & & & & &\\
& & &  & & & & & &\\
HS1232+3947 & -17.16 & 0.0210 & 29 & 0.286 & 1.09e+40 & 0.25 & -1.87 & V & DHIIH \\          
HS1236+3821 & -17.06 & 0.0073 &  4 & 0.602 & 2.36e+39 & 0.05 & -2.66 & V & DHIIH \\     
HS1236+3937 & -15.67 & 0.0184 &112 & 0.081 & 3.35e+39 & 0.08 & -2.03 & V & DHIIH/SS \\               
HS1256+3505 & -18.75 & 0.0342 & 20 & 0.448 & 7.54e+40 & 0.71 & -1.61 & F & DANS  \\           
HS1258+3438 & -16.4  & 0.0248 & 36 & 0.380 & 9.31e+39 & 0.21 &   *   & F & DHIIH \\ 
HS1301+3325 & -16.96 & 0.0246 & 11 & 0.573 & 1.21e+40 & 0.27 & -1.74 & F & DHIIH \\             
HS1301+3209 & -17.06 & 0.0238 & 10 & 0.579 & 9.29e+39 & 0.21 & -1.94 & F & HIIH \\            
HS1304+3529 & -17.3  & 0.0165 &118 & 0.013 & 1.94e+40 & 0.44 & -1.72 & F & IP \\          
HS1306+3320 & -18.32 & 0.0270 & 36 & 0.508 & 4.45e+40 & 1.01 & -1.87 & V & HIIH \\        
HS1308+3044 & -17.92 & 0.0209 &  4 & 0.564 & 1.05e+40 & 0.24 & -2.13 & F & DANS \\            
HS1311+3628 & -16.84 & 0.0031 &301 & 0.160 &     *    &    * &   *   & F & DHIIH \\        
HS1312+3508 & -16.26 & 0.0035 &254 & 0.164 & 2.62e+39 & 0.06 & -2.33 & F & DHIIH \\       
HS1315+3132 & -16.46 & 0.0315 & 36 & 0.476 & 1.77e+40 & 0.40 & -1.45 & F & DHIIH \\     
HS1319+3224 & -15.3  & 0.0182 & 49 & 0.208 &     *    &    * &   *   & F & SS/DHIIH \\   
HS1325+3255 & -15.92 & 0.0263 & 74 & 0.146 & 6.09e+39 & 0.14 & -1.71 & V & DHIIH/SS \\
HS1328+3424 & -17.79 & 0.0227 &  8 & 0.503 & 5.07e+39 & 0.11 & -2.60 & V & HIIH  \\                  
HS1330+3651 & -17.15 & 0.0167 & 72 & 0.147 & 1.47e+40 & 0.33 & -1.84 & F & DHIIH \\         
HS1332+3426 & -16.14 & 0.0220 & 35 & 0.444 & 8.25e+39 & 0.19 & -1.74 & V & DHIIH \\       
HS1334+3957 & -15.32 & 0.0083 & 71 & 0.000 &     *    &    * &    *  & F & DHIIH \\   
HS1336+3114 & -17.94 & 0.0158 &  5 & 0.000 &     *    &    * &    *  & F & HIIH  \\      
HS1340+3307 & -16.63 & 0.0158 & 20 & 0.539 & 1.21e+40 & 0.27 & -1.50 & F & DHIIH \\      
HS1341+3409 & -16.35 & 0.0171 & 15 & 0.480 & 6.77e+39 & 0.15 & -1.77 & F & DHIIH \\      
HS1347+3811 & -15.30 & 0.0103 & 64 & 0.243 & 2.30e+39 & 0.05 & -2.01 & F & DHIIH \\     
HS1349+3942 & -15.16 & 0.0054 & 13 & 0.337 & 9.56e+38 & 0.02 & -2.16 & V & DHIIH \\    
HS1354+3634 & -17.10 & 0.0167 & 19 & 0.513 & 1.46e+40 & 0.33 & -1.74 & F & DANS \\         
HS1354+3635 & -17.81 & 0.0171 & 21 & 0.507 & 2.75e+40 & 0.62 & -1.57 & F & HIIH \\          
HS1402+3650 & -18.81 & 0.0347 & 26 & 0.561 & 8.56e+40 & 1.94 & -1.69 & F & HIIH \\        
HS1410+3627 & -18.11 & 0.0338 & 15 & 0.311 & 1.73e+40 & 0.39 & -2.11 & F & HIIH \\        
HS1416+3554 & -16.94 & 0.0103 & 14 & 0.328 & 2.50e+39 & 0.06 & -2.61 & F & DHIIH/HIIH \\       
HS1420+3437 & -16.75 & 0.0246 & 27 & 0.448 & 1.84e+40 & 0.42 & -1.39 & F & DHIIH \\
HS1422+3325 & -17.30 & 0.0341 & 20 & 0.337 & 1.70e+40 & 0.39 & -1.71 & F & HIIH \\             
HS1422+3339 & -16.4  & 0.0114 & 17 & 0.350 & 4.68e+39 & 0.11 &    *  & F & DHIIH/HIIH \\       
HS1424+3836 & -16.06 & 0.0218 &104 & 0.195 & 1.07e+40 & 0.24 & -1.47 & F & DHIIH \\
HS1425+3835 & -17.80 & 0.0223 & 12 & 0.099 &     *    &    * &    *  & F & HIIH \\             
HS1429+4511 & -17.40 & 0.0321 & 15 & 0.307 &     *    &    * &    *  & V & DANS \\                 
HS1429+3154 & -16.73 & 0.0117 & 34 & 0.143 & 5.63e+39 & 0.13 & -2.04 & F & DHIIH \\
HS1440+4302 & -15.07 & 0.0085 & 44 & 0.336 & 2.54e+39 & 0.06 & -1.82 & F & DHIIH/SS \\          
HS1440+3120 & -17.4  & 0.0525 &152 & 0.221 & 6.32e+40 & 1.43 &    *  & F & DHIIH \\
HS1440+3805 & -18.80 & 0.0322 &  7 & 0.263 &     *    &    * &    *  & F & HIIH \\
HS1442+4250 & -14.73 & 0.0025 &113 & 0.081 & 3.19e+38 & 0.01 & -2.64 & F & SS \\              
HS1444+3114 & -19.02 & 0.0297 & 26 & 0.413 & 9.12e+40 & 2.06 & -1.68 & F & DANS \\
HS1502+4152 & -16.66 & 0.0164 &  7 & 0.515 & 2.81e+39 & 0.06 & -2.52 & F & DHIIH \\
HS1507+3743 & -17.58 & 0.0322 &232 & 0.114 & 6.68e+40 & 1.51 & -1.21 & V & DHIIH \\
HS1529+4512 & -17.05 & 0.0231 & 16 & 0.222 & 3.77e+39 & 0.09 & -2.46 & V & HIIH/DHIIH \\           
HS1544+4736 & -17.11 & 0.0195 & 32 & 0.025 & 5.79e+39 & 0.13 & -1.98 & F & DHIIH \\
HS1609+4827 & -17.6  & 0.0096 & 12 & 0.381 & 6.08e+39 & 0.14 &    *  & F & HIIH \\           
HS1610+4539 & -17.2  & 0.0196 & 26 & 0.480 & 2.35e+40 & 0.53 &    *  & F & DHIIH/HIIH \\        
HS1614+4709 & -13.6  & 0.0026 &140 & 0.100 & 8.09e+38 & 0.02 &    *  & F & SS \\
HS1633+4703 & -15.93 & 0.0086 & 15 & 0.364 & 3.29e+39 & 0.07 & -1.98 & F & DHIIH \\                  
HS1640+5136 & -19.59 & 0.0308 & 21 & 0.443 & 9.15e+40 & 2.07 & -1.71 & F & SBN/HIIH \\                   
HS1641+5053 & -19.05 & 0.0292 & 15 & 0.574 & 9.69e+40 & 2.19 & -1.67 & F & HIIH \\             
HS1645+5155 & -19.11 & 0.0286 & 47 & 0.330 & 3.84e+40 & 0.87 & -2.21 & F & HIIH \\            
HS1723+5631a & -18.70 & 0.0286 & 23 & 0.341 & 2.91e+40 & 0.66 & -1.88 & F & IP \\              
HS1723+5631b & -18.70 & 0.0286 & 32 & 0.337 & 2.96e+40 & 0.67 & -1.88 & F & IP \\               
HS1728+5655 & -16.38 & 0.0160 &106 & 0.094 & 1.09e+40 & 0.25 & -1.46 & F & DHIIH \\            
\hline
\end{tabular}

\end{table*}

Following again Kennicutt
(1983) we derive the ratio of the current SFR to the average past SFR, 
${\rm b} = {\rm SFR}/<{\rm SFR}>$, the
\begin{figure}[htp]
\includegraphics[scale=0.4]{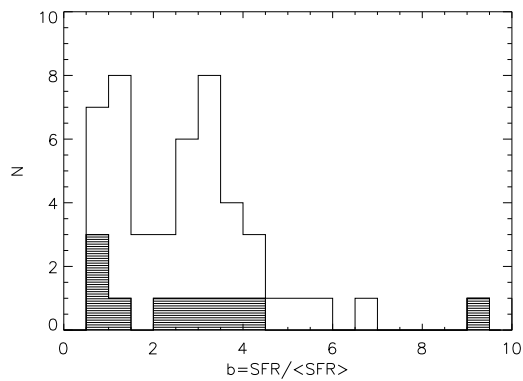}
\caption{The distribution of the b ratio between the current SFR and the 
averaged past SFR. The solid histogram is the distribution of the whole sample,
while the hashed histogram is for Void galaxies.}
\end{figure}
latter being derived as the mass of an underlying older disk population, M$_d$,
divided to its age, fixed to 10\,Gyr. The disk mass was obtained by 
multiplying the blue
luminosity of the galaxy with an average mass-to-luminosity ratio of
1\,M$_{\odot}/{\rm L}_{B{\odot}}$. The M/L$_B$ was taken from Kennicutt,
Tamblyn and Congdon (1994) (see their Table\,\,3 for the irregular class of
galaxies) and it is adopted as an 
average value for the whole sample of BCDs. Nevertheless, some of the 
galaxies with strong stellar bursts (larger W(H${\beta})$ values) may have 
significantly lower mass-to-luminosity ratios, 
and thus a larger value for the b Scalo parameter. Therefore our b values are 
only lower limits. While the absolute calibration of the b parameter may be
subject of considerable uncertainties (up to a factor of 2), this should not 
affect the comparison between the b distribution of the Field and Void 
samples. The result of the 
calculations (Fig. 5) shows  that most of the galaxies are dominated
by recent star formation episodes, and there is no trend with the environment.
The galaxy with the highest current star-formation rate with
respect to the averaged past (a factor 9) is in fact a Void galaxy, 
HS1507+3743.

\begin{figure}[htp]
\includegraphics[scale=0.4]{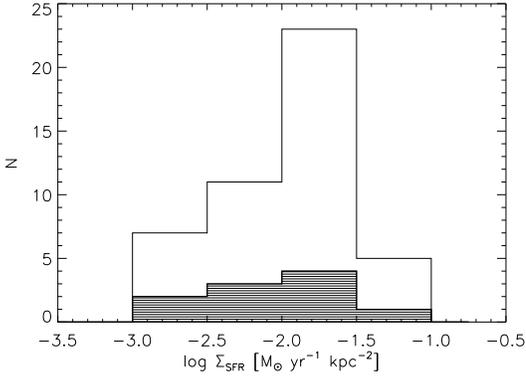}
\caption{The distribution of the mean SFR surface density ${\Sigma}_{SFR}$. 
The solid histogram is the distribution of the whole sample, while the hashed 
histogram is for Void galaxies.}
\end{figure} 
 
Finally, the mean SFR surface density ${\Sigma}_{SFR}$
(M$_{\odot}\,{\rm yr}^{-1}$\, ${\rm kpc}^{-2}$) was derived for each galaxy by
dividing the total SFR from equation (1) by the area within the isophotal 
r$_{25}$ 
radius (in kpc) (Vennik et al. (1999)). The derived ${\Sigma}_{SFR}$ are listed
in Table 1, Column 8 and their distributions are shown in Fig. 6. The histogram
of Void galaxies follows that of the Field galaxies.

\section{Diagnostic diagrams and metallicities}

It has been suggested (V\'{\i}lchez 1995) that galaxies showing HII region
spectra of very high excitation and evidence for hard ionizing spectra, as
indicated by the ratio of the
\newline
[OIII]\,${\lambda}$5007,4959/[OII]\,${\lambda}$3727, are associated with the
low density regions. 
\begin{figure}[htp]
\includegraphics[scale=0.4]{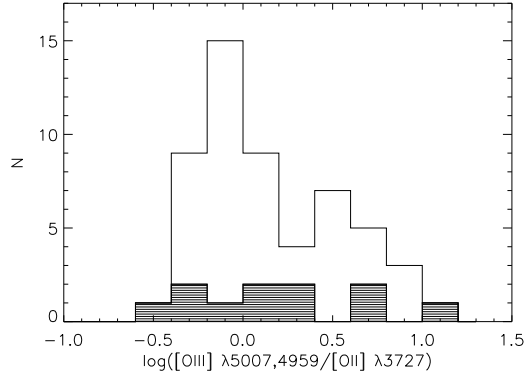}
\caption{The distribution of the 
log([OIII]\,${\lambda}$5007,4959/ [OII]\,${\lambda}$3727) ratio. The solid 
histogram is the distribution of the whole sample, while the hashed histogram 
is for Void galaxies.}
\end{figure}
In Fig. 7 we show that the distribution of the [OIII]/[OII] is not different
between Field and Void galaxies, the same being true for the distribution of
the equivalent widths of the [OIII]$\lambda$5007 (Fig. 8).

\begin{figure}[htp]
\includegraphics[scale=0.4]{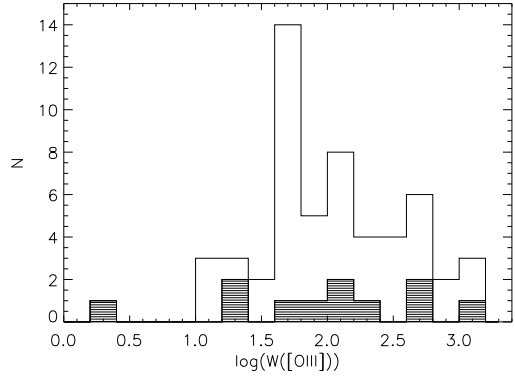}
\caption{The distribution of the [OIII]\,${\lambda}$5007 equivalent widths (in
units of ${\AA}$). 
The solid histogram is the distribution of the whole sample,
while the hashed histogram is for Void galaxies.}
\end{figure}

\begin{figure}[htp]
\includegraphics[scale=0.4]{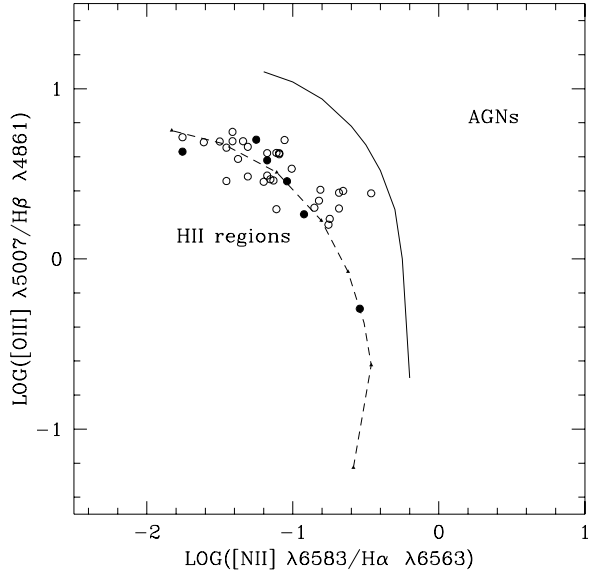}
\caption{Line diagnostic of log([OIII]\,${\lambda}5007$/H${\beta}$) vs
log([NII\,${\lambda}6583$/H${\alpha}$). Open circles are Field galaxies and 
filled circles are Void galaxies. The dotted curve represents the HII region
models of Dopita and Evans (1986), calibrated in metallicity according to
Salzer et al. (1989b). The solid triangles located along the curve indicate
different values for the metallicity of the gas: 2.0\,Z${\odot}$ at the lower
right-hand end of the curve, with successive symbols representing
1.5, 1.0, 0.75, 0.50, and 0.25 times solar, respectively. The solid line 
divides AGNs from HII region-like objects.} 
\end{figure}

\begin{figure}[htp]
\includegraphics[scale=0.4]{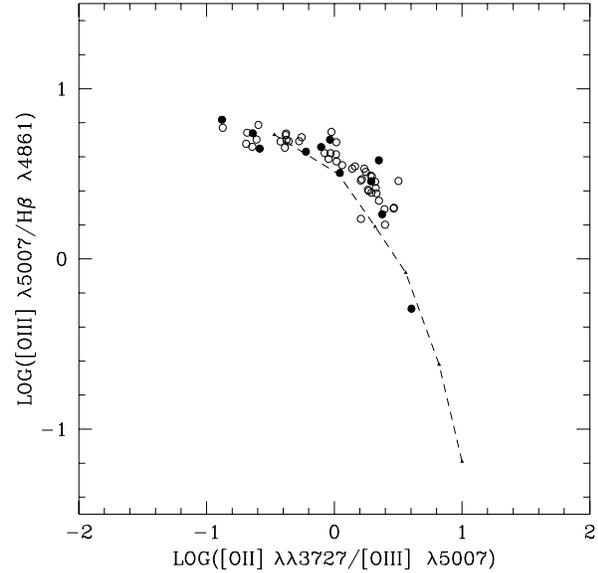}
\caption{Line diagnostic of log([OIII]\,${\lambda}5007$/H${\beta}$) vs
log([OII]\,${\lambda}$3727/[OIII]\,${\lambda}5007$). Open circles are Field 
galaxies and filled circles are Void galaxies. The symbols like in Fig. 9.} 
\end{figure}

Because accurate measurements of the electron temperature are not available for
most of our galaxies, we used diagnostic diagrams (Veilleux and Osterbrock
1987) and the models of Dopita and
Evans (1986) in order to obtain oxygen abundances. In Fig. 9 and 10 we plotted  
the diagnostic diagrams [OIII]/H${\beta}$ versus
[NII]\,${\lambda}6583$/H${\alpha}$ and [OIII]/H${\beta}$ versus 
[OII]\,${\lambda}$3727/[OIII]\,${\lambda}5007$. Our data seem to follow quite
well the models of Dopita and Evans (1986) (shown as dotted lines
in the diagnostic diagrams), suggesting that our
objects are typical HII galaxies, most of them of high to intermediate
excitation, and thus of low to intermediate metallicities. Following Salzer et
al. (1989b) we have calibrated the HII model from the diagnostic diagrams in
metallicity: from upper left to lower right, 0.25, 0.50, 0.75, 1.0, 1.5 and 2.0
times solar metal abundance (we consider a solar metal abundance 12+log(O/H) =
8.73)). We obtained the metallicity of each object
separately, from each diagnostic diagram, and the assigned metallicity was given
as an averaged of the two independent estimates. The differences between the
two methods are smaller than $\Delta$ (12+log(O/H)) = 0.1 dex. We should mention that the 
zero point of this calibration is not known, but since we are interested only 
in a relative comparison between Field and Void galaxies, the knowledge of 
the absolute metallicity is not required. 

\begin{figure}[htp]
\includegraphics[scale=0.4]{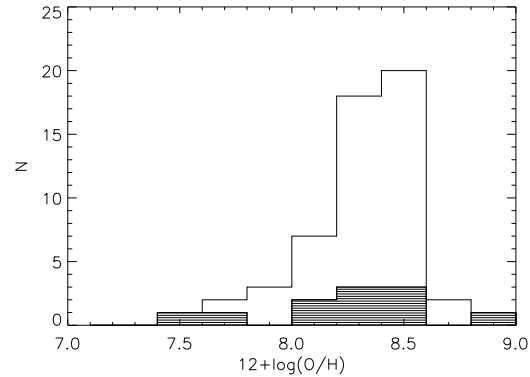}
\caption{The distribution of the oxygen abundances 12+log(O/H). The solid
histogram is the distribution for the whole sample, while the hashed histogram
is for Void galaxies.} 
\end{figure} 

For a few galaxies the temperature sensitive line \newline 
[OIII]~${\lambda}$4363 was
detected.  Our spectra are of low spectral resolution and therefore even for
high temperature HII regions it is difficult to obtain an accurate flux value
for this line in the blend with the much stronger H${\gamma}$ line. We used a
five level atom program and the ionization correction method of Mathis and 
Rosa (1991) to obtain O/H abundances. All of these are lower than 0.25 solar, a
region where the strong line methods, i.e. of [OIII]/[OII] do not yield O/H
unambiguously and/or are rather insensitive to the actual value of O/H. In
conclusion we can use the O/H determinations only as an indication of the
magnitude of the metallicities, indicated by binning in 0.5 dex intervals in
Fig. 11. Nevertheless, Fig. 11 shows again that both the Field and the Void
population occupy the same parameter space in metallicity range.

\section{The Field - Cluster dichotomy?}

Our study has shown that the dwarf Void galaxies are very similar in their star
formation histories and metallicities to Field galaxies of the same
morphological type. It appears as if the Void galaxies are
unaware of the fact that they exist in huge underdense regions. Going to
the opposite environment, we would like to know whether the BCDs in the core
regions of rich clusters may show different properties. Our data do not cover
any rich cluster, therefore we cannot directly address this question, but we
investigated available samples in the literature. 

V\'{\i}lchez (1995)
studied the star-forming galaxies in the core of the Virgo Cluster with respect
to star-forming galaxies from low density environments. He found a clear trend,
with ELGs from low-density environments presenting higher excitation and
ionization parameters, higher H${\beta}$ equivalent widths, and larger total 
H${\beta}$ luminosities than objects located in the Virgo Cluster
core. He interprets these results as being consistent with the existence of
two populations of dwarfs, with different origins and evolutionary
histories: The BCDs from low density environments experience their first 
big episodes of 
star formations, while the Virgo cluster 
star-forming dwarfs have a mixed star-formation history, requiring 
continuous star formation in addition to some currently observable bursts. 

However, the sample of Virgo cluster core was
selected from the Virgo Cluster Catalogue (VCC) of Binggeli et al. (1989), which
was based on a visual classification of the direct images, while the
dwarfs from low-density regions were taken from the University of Michigan 
objective-prism survey (MacAlpine and Lewis 1978, MacAlpine and Williams 1981, 
Salzer 1989). It is well known that objective-prism surveys that use 
IIIa-J plates are very sensitive in detecting high ionization HII galaxies, 
since the selection criteria is based on the presence of the 
[OIII]\,${\lambda}$5007 emission
line. These kind of surveys can select intrinsically very faint objects,
very compact, with almost stellar-like appearance and almost no underlying
older stellar population. Such extreme objects can be very easily missed from
surveys that select objects based on morphological criteria, or discriminate
galaxies from stars based on the apparent diameter of the latter. On the other
hand IIIa-J objective prism surveys are less sensitive on lower ionization
objects, with stronger H${\alpha}$ lines and stronger continuum, which are 
better detected from direct plates (or H${\alpha}$ surveys). Therefore we 
suggest that the dichotomy found by V\'{\i}lchez between the spectroscopic 
properties of Virgo Cluster BCDs and star forming galaxies from lower density 
regions reflect mainly the differences between morphological types, namely 
between extreme SS/DHIIH galaxies and HIIH/DANS/Im/GI (giant irregular). 
V\'{\i}lchez (1995) argued that 
the BCD and Sm/BCD or Im/BCD classes found in the Virgo core are equivalent 
with the Salzer HII galaxies. 

In order to prove that the BCDs from the Virgo
Cluster core are biased towards subtypes of  intermediate properties, we have 
reclassified these objects according to the Salzer classes, based on their 
morphologies, absolute magnitudes and diameters. We should mention that we 
did not find any extreme SS type among the Virgo sample. The galaxy VCC1437 
from the sample of V\'{\i}lchez (1995) was not included, since it is considered
 an Elliptical galaxy in NED. In Fig. 12 we present the
\begin{figure}[htp]
\includegraphics[scale=0.4]{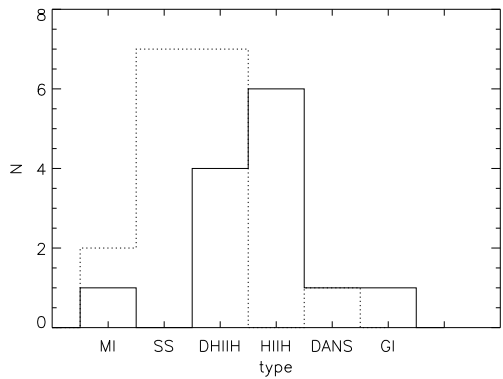}
\caption{The distribution of morphological types for the galaxies in the Virgo
cluster core (solid histogram) and for the galaxies in the Virgo cluster
periphery (dotted histogram).} 
\end{figure} 
distribution of morphological types for the Virgo sample (solid histogram), as
compared to the distribution of morphological types for the Virgo cluster 
periphery. The two distributions are clearly different. The Virgo cluster 
periphery is dominated by SS and DHIIH galaxies, which consist of very high 
ionization objects with very recent star bursts, which dominate over the older 
stellar population. On the other hand the Virgo cluster core is dominated by
HIIH galaxies. We conclude that the two samples were not suitable for studying
environmental effects since they contain different objects. Nevertheless,
it is still possible that the Virgo cluster core is
deficient in extreme BCDs of SS/DHIIH type. In order to draw a definitive 
conclusion one should perform a dedicated search towards the Virgo cluster 
core, using the same selection criteria as for the Virgo periphery or field
regions.  We conclude that with the presently available data it is not 
possible to investigate the influence of the cluster environment on the 
star-formation properties of BCDs and hence also on their metallicities.

\section{Discussion}

In the decade following the discovery of BCDs, several investigations based on
different samples showed that most of them are metal poor systems (Searle and
Sargent 1972, Lequeux et al. 1979, French 1980, Kunth and Sargent 1983). It was
thought that some of these galaxies are young systems in the process of forming
their first generation of stars. But
the detection of an extended faint stellar underlying component in the
majority of BCDs (Loose and Thuan 1986, Kunth et
al. 1988, Telles, Melnick, and Terlevich (1997), Vennik et al. 1999) 
supports the idea that they are not truly 
primordial galaxies, but older LSB dwarf galaxies undergoing transient 
periods of star formation. A clear demonstration of this two-component 
structure was recently presented by Schulte-Ladbeck et al. (1998, 1999) for 
the very nearby BCD VII Zw 403, based on HST photometry of its individual 
stars down to the red giant branch stars. 

Papaderos et al. (1996) proposed a scenario in which the BCDs 
consist of an old ($\ge 10$ Gyr) stellar component embedded within a larger 
self-gravitating
gaseous envelope. In this scenario slow infall to the center may produce a
central gas density large enough to initiate the first starburst event. A few
Myr after the formation of the first massive stellar population, multiple SNe
explosions will prevent the cold gas from further collapse. Depending on the
mass of the BCD, galactic winds can produce different mass losses of gas and
metals. Subsequent infall of the outer HI gas will replenish the cavity in the
ISM formed during the previous starburst and, after a quiescent period
determined by the gas cooling and mass accretion timescales, the gas density
threshold for star formation may be reached again. If we look on the sequence
of BCD types, from SS to DANS, we can imagine that smaller LSB will evolve 
into SS and DHIIH classes, while the bigger ones into HIIH and DANS classes. 
Since smaller systems have lower gravitational potentials, galactic winds can 
lead to a significant loss of metals, and thus to the formation of very low
metallicity systems. 

In the core of rich clusters, the outer gas envelopes of these 
LSB dwarf may be stripped, leaving the galaxies without their gas reservoir. 
Then these objects will not have the chance to initiate star-bursts and they
will thus not change into BCDs. The smallest systems should be most affected, 
which may result of a depletion
of SS and DHIIH in the core of rich clusters. The results of Vilchez for the 
star-forming galaxies
in the Virgo cluster core can then be explained in these terms. Further, the
pressure of the intracluster medium should prevent significant mass loss due to
galactic winds, and the surviving BCDs should have higher metallicities than
their counterparts in the field. As mentioned in the previous section, until 
we possess a dedicated, bias free search towards the Virgo cluster core, the
discussion above remains very speculative.

In the other extreme environments, the voids, we expect galaxies to have larger
gas reservoirs than in field, because here there are no destruction 
mechanisms. Nevertheless, if the infall of gas is a slow process, producing 
star bursts in short
episodes, then the star-formation properties of these galaxies should not be
different from those of field galaxies, as concluded by our study. They may
only have enough gas to supply star-bursts for a longer time than in the field,
but at the present epoch their properties should not be different.

Recently Lemson and Kauffmann (1999) have considered the environmental 
influences on dark matter halos and therefore the consequences for the galaxies
within them. They use large N-body simulations of dissipationless gravitational
clustering in cold dark matter (CDM) cosmologies and they found that the mass
of the dark matter halo is the only halo property that correlates 
significantly with the local environment. The dependence of galaxy 
morphology, luminosity, surface brightness and star formation rate on 
environment, must all arise because galaxies are preferentially found in 
higher mass halos in overdense environments and in lower mass halos in 
underdense environments. On the other hand, there seem to be little 
difference between the mass function of dark matter halos for 
field and void environments. 
If the models of Lemson and
Kauffmann are correct, and indeed there is no correlation between the formation
redshifts, concentrations, shapes and spins of the dark matter halos and the
overdensity of their local environment, then the star-formation properties of 
BCDs in Voids and Field are expected to be similar, at least in the frame of
the current hierarchical CDM cosmological models.

In summary, we found that the star formation rates and metallicities of BCDs 
do not show any significant dependency on environment between the densities
encountered in Voids and in the Field. For high-density environments there 
could be a lack of extreme BCDs of SS/DHIIH type, and therefore a difference 
in the star-formation properties of cluster BCDs, as suggested by 
V\'{\i}lchez (1995) for the Virgo Cluster star-forming galaxies. We argue 
that available samples in the core of rich clusters are not suitable for this 
kind of analysis.

\begin{acknowledgements}

We would like to thank Dr. Jaan Vennik for performing the morphological 
classification of the star - forming galaxies in the Virgo Cluster core.
We gratefully  acknowledge Profs. H. Els\"asser and B. Binggeli for 
interesting discussions. U.H. acknowledges the support of the SFB 375.

This research has made use of the NASA/IPAC Extragalactic Database (NED) which
is operated by the Jet Propulsion Laboratory, California Institute of
Technology, under contract with the National Aeronautics and Space 
Administration.

\end{acknowledgements}

\end{document}